\documentstyle [12pt] {article}
\begin{document}
\def\be{\begin{eqnarray}}
\def\en{\end{eqnarray}}
\def\non{\nonumber}
\def\la{\langle}
\def\ra{\rangle}
\def\ep{\varepsilon}
\def\pr{{\sl Phys. Rev.}~}
\def\prl{{\sl Phys. Rev. Lett.}~}
\def\pl{{\sl Phys. Lett.}~}
\def\np{{\sl Nucl. Phys.}~}
\def\zp{{\sl Z. Phys.}~}

\font\el=cmbx10 scaled \magstep2
{\obeylines
\hfill IP-ASTP-11-94
\hfill June, 1994}

\vskip 1.5 cm

\centerline{\el Nonfactorizable Contributions to Nonleptonic Weak Decays}
\centerline{\el of Heavy Mesons}
\medskip
\bigskip
\medskip
\centerline{\bf Hai-Yang Cheng}
\medskip
\centerline{ Institute of Physics, Academia Sinica}
\centerline{Taipei, Taiwan 11529, Republic of China}
\bigskip
\bigskip
\bigskip
\centerline{\bf Abstract}
\bigskip
   Nonfactorizable contributions to exclusive two-body nonleptonic weak decays
of heavy mesons arising from color octet currents, characterized by the
parameter $r_2$, are extracted from the data. It is found that $r_2$ is
equal to $\sim -0.67\,$, $-(0.9\sim 1.1),~-(1.2\sim 1.3)$, $\sim 0.36$
respectively for $D\to \bar{K}\pi,~\bar{K}^*\pi,~\bar{K}^*\rho,~\bar{B}\to
D\pi$ decays. As expected, soft-gluon effects become stronger when the
decay particles are less energetic, allowing more time for significant
final-state strong interactions. As a consequence, the parameter $a_2$ is not
universal and is channel or class dependent, contrary to the
anticipation of the factorization approach. The leading $1/N_c$ expansion
works most successfully for $D\to\bar{K}\pi$ decays as the subleading
$1/N_c$ factorizable contribution is almost compensated by the soft-gluon
effect. We argue that, in contrast to what happens in $B^-\to D^{(*)}
\pi(\rho)$ decays, the nonfactorizable term and the subleading $1/N_c$
factorizable term in $\bar{B}\to \psi\bar{K}^{(*)}$ decays are opposite in
signs, in accordance with a recent QCD sum rule calculation.
Implications are discussed.
\pagebreak

    1. {\it Introduction}~~~A common wisdom for having an approximate
description of
the underlying physics of nonperturbative QCD is to search for the small
parameters in the theory, then set the small parameters to zero and treat the
finite effects of the parameters as small perturbations [1,2]. Hence, even
for a nonperturbative physics such as QCD, one can still perform a sensible
perturbative calculation. The well-known examples are chiral perturbation
theory and the heavy quark effective theory which respresent a good
approximation of QCD in the light and heavy quark mass limits, respectively.
The corresponding small expansion parameters in these theories are
$m_q/\Lambda_{\rm QCD}$ and $\Lambda_{\rm QCD}/m_Q$ with $m_q~(m_Q)$ being
the light (heavy) quark mass.

     By treating the number of color degrees of freedom as a free parameter
$N_c$, $1/N_c$ can be another useful small parameter in QCD. The fact that
$1/N_c$ is indepedent of energy has its advantage and disadvantage. On the
one hand, it can be used as an expansion parameter for both short- and
long-distance dynamics owing to its energy independence. On the other hand,
whether or not the $1/N_c$ approach works is at best case by case dependent,
since, unlike chiral perturbation theory or the heavy quark effective theory,
there is no certain kinematic region where the validity of the $1/N_c$
expansion is guaranteed. Empirically, while the large-$N_c$ approach fails
in some cases, it explains well the
OZI rule and even provides quantitative predictions on the relative
strength of the baryon-meson coupling constants and on the baryon mass
relations up to ${\cal O}(1/N_c^2)$, as was realized last year [3]. It is also
known for sometime that the leading $1/N_c$ expansion operates reasonably well
for exclusive two-body nonleptonic decays of charmed mesons [4];
\footnote{For a review of the $1/N_c$ expansion for nonleptonic weak decays
of mesons, see Ref.[5].}
 the discrepancy
between theory and experiment for color-suppressed channels e.g. $D^0\to
\bar{K}^0\pi^0$ is greatly improved provided that contributions from
Fierz-transformed currents, which are suppressed by order $1/N_c$, are
dropped [6,7]. Due to the success of the $1/N_c$ approach to charmed meson
decays, it has been widely believed by many practitioners in this field that
it applies equally well to the weak decays of bottom mesons.

     The recent CLEO data on the $B$ decays $\bar{B}\to D\pi,~D\rho,~D^*\pi,~
D^*\rho$ exhibit a quite striking result [8]: The interference between the two
different amplitudes contributing to exclusive two-body $B^-$ decays are
evidently constructive, contrary to what naively expected from the leading
$1/N_c$ expansion. In other words, the observed destructive interference
pattern in $D^+\to \bar{K}^0\pi^+,~\bar{K}^0\rho^+,~\bar{K}^{*0}\pi^+$ cannot
be extrapolated to $B$ decays.
Though, as stressed before, the rule of retaining only the leading term in the
$1/N_c$ expansion is case by case dependent,
an understanding of why it is operative for charm decays but not
for the $B$ meson case is called for.

     Since $N_c=3$ in reality, when the $1/N_c$ approach works empirically,
it will imply either that the effective expansion parameter is something like
\footnote{For example, the expansion parameter in QED is $\alpha=e^2/4\pi$
even though $e\sim {1\over 3}$ is not very small.}
$1/(4\pi N_c),~1/N_c^2$..., or that there is a dynamic reason for the
suppression of non-leading $1/N_c$ terms. In the large-$N_c$ limit, the meson
decay amplitude is factorizable [4] (this is no longer true for baryon decay
as the baryon contains $N_c$ quarks.) Irrespective of the $1/N_c$ expansion,
the factorization hypothesis means that the meson two-body decay amplitude
may be expressed as the product of two independent hadronic currents.
Consider the operator
\be
O_1=\,(\bar{s}c)(\bar{u}d),
\en
where $(\bar{q}_1q_2)=\bar{q}_1\gamma_\mu(1-\gamma_5)q_2$. The decay amplitude
of $D^+\to \bar{K}^0\pi^+$ induced by $O_1$ in general can be expressed
in terms of factorizable and nonfactorizable contributions:
\be
\la\bar{K}^0\pi^+|O_1|D^+\ra= \,\la\pi^+|(\bar{u}d)|0\ra\la\bar{K}^0|(\bar
{s}c)|D^+\ra+\la\bar{K}^0\pi^+|O_1|D^+\ra_{nf},
\en
where the subscript $nf$ denotes a nonfactorizable contribution. In the
$1/N_c$ expansion, the factorizable amplitude is of order $N_c^{1/2}$, while
the  nonfactorizable one is of order $N_c^{-1/2}$. Using the identity
\be
O_1 =\, {1\over N_c}O_2+{1\over 2}(\bar{s}\lambda^a d)(\bar{u}\lambda^a c),
\en
with  $O_2 = (\bar{s}d)(\bar{u}c)$ and $(\bar{q}_1\lambda^aq_2)=\bar{q}_1
\gamma_\mu(1-\gamma_5)\lambda^aq_2$, one obtains
\be
\la\bar{K}^0\pi^+|O_1|D^+\ra_{nf} &=& {1\over N_c}\la\bar{K}^0\pi^+|O_2|D^+\ra
_f+{1\over 2}\la\bar{K}^0\pi^+|(\bar{s}\lambda^a d)(\bar{u}\lambda^a c)|D^+
\ra   \non \\     &+&{1\over N_c}\la\bar{K}^0\pi^+|O_2|D^+\ra_{nf},
\en
where the factorizable amplitude $\la\bar{K}^0\pi^+|O_2|D^+\ra_f$ is equal to
$\la\bar{K}^0|(\bar{s}d)|0\ra\la\pi^+|(\bar{u}c)|D^+\ra$. In the traditional
vacuum insertion method only the factorizable terms are retained so that
\be
\la\bar{K}^0\pi^+|O_1|D^+\ra_{nf}=\,{1\over N_c}\la\bar{K}^0\pi^+|O_1|D^+\ra
_{f}.
\en
However, this naive method encounters two problems. First, it is not
logically consistent [9]: A nonfactorizable term becomes a subleading
$1/N_c$ factorizable contribution.
Second, naive factorization cannot explain the bulk of the experimental
data of charm decay (for a review, see Ref.[5]). Both problems hint that
it is indispensible to take into account nonfactorizable effects. To the order
${\cal O}(N_c^{-1/2})$, we thus keep the second nonfactorizable term on the
r.h.s. of Eq.(4) and drop the third term as it is further suppressed by a
factor of $1/N_c$. To the end, we have
\be
\la\bar{K}^0\pi^+|O_1|D^+\ra=\,\la\bar{K}^0\pi^+|O_1|D^+\ra_{f}+
\la\bar{K}^0\pi^+|O_2|D^+\ra_{f}\left({1\over N_c}+{r_2\over 2}\right),
\en
where
\be
r_2=\,{\la\bar{K}^0\pi^+|(\bar{s}\lambda^a d)(\bar{u}\lambda^a c)|D^+\ra
\over \la\bar{K}^0\pi^+|(\bar{s} d)(\bar{u}c)|D^+\ra_f},
\en
first introduced in Ref.[10],
denotes the nonperturbative effects arising from the soft gluon exchange
between the color octet currents $(\bar{s}\lambda^a d)$ and
$(\bar{u}\lambda^a c)$ relative to that from the corresponding color singlet
currents and is of order $1/N_c$ in the large-$N_c$ limit.

   It is clear from Eq.(6) that if the leading $1/N_c$ expansion works, it
will come from the dynamical reason that the subleading $1/N_c$ factorizable
contribution is compensated by the nonperturbative correction, namely
$r\approx -2/N_c$. Though in practice it is very difficult to estimate the
nonperturbative soft gluon effects (for some recent attempts; see
Refs.[11-13]), they can be extracted from the available data. Consider the
decays $D\to PP,~VP,~VV$ and $B\to PP$ ($P$: pseudoscalar meson, $V$: vector
meson). It is expected that the soft-gluon effect is such that
\be
|r_2(D\to VV)|>|r_2(D\to VP)|>|r_2(D\to PP)|>|r_2(B\to PP)|,
\en
as the final-state particles have more time to allow significant final-state
strong interactions when they become less energetic.
The purpose of this paper is to determine the parameter
$r_2$ from data and confirm the above expectation. Implications on
the factorization method are discussed.
\vskip 1.0 cm
2.~~{\it Charmed Meson Decays}~~~
In the large-$N_c$ approach both soft-gluon nonperturbative effects
and final-state interactions are subleading $1/N_c$ nonfactorizable
corrections [4]. In order to determine the soft-gluon effects, we will focus
on the exotic channels e.g. $D^+\to\bar{K}^0\pi^+,~\pi^0\pi^+$ and the
decay modes with one single isospin component, e.g. $D^+\to\pi^+\phi,~D_s^+
\to\pi^+\phi$, where final-state interactions are presumably negligible.

    The QCD-corrected weak Hamiltonian for Cabibbo-allowed charm decays is
given by
\be
{\cal H}_{\rm eff}=\,{G_F\over\sqrt{2}}V_{ud}^*V_{cs}(c_1O_1+c_2O_2),
\en
with $O_1=(\bar{s}c)(\bar{u}d),~O_2=(\bar{s}d)(\bar{u}c)$, $c_1$ and $c_2$ are
Wilson coefficient functions determined at the renormalization scale
$\mu\sim m_c$
\be
c_1(m_c)\sim 1.26\,,~~~~c_2(m_c)\sim -0.51\,.
\en
The amplitude for the decay $D^+\to\bar{K}^0\pi^+$ up to the subleading
order in $1/N_c$ is
\be
&& A(D^+\to\bar{K}^0\pi^+) =\,{G_F\over\sqrt{2}}V_{ud}^*V_{cs}(a+b), \non \\
&& a =\, a_1(m_D^2-m_K^2)f_\pi f_+^{DK}(m_\pi^2), \\
&& b =\, a_2(m_D^2-m_\pi^2)f_Kf_+^{D\pi}(m_K^2), \non
\en
where $f_\pi=132$ MeV, $f_K=160$ MeV,
\be
a_1=\,c_1+\xi_1c_2,~~~~a_2=\,c_2+\xi_2c_1,
\en
with $\xi_1=1/N_c+r_1/ 2,~\xi_2=1/N_c+r_2/ 2$,
and
\be
r_1(D\to\bar{K}\pi) &=& {\la\pi^+\bar{K}^0|(\bar{u}\lambda^a d)(\bar{s}
\lambda^a c)|D^+\ra \over \la\pi^+\bar{K}^0|(\bar{u}d)(\bar{s}c)|D^+\ra_f}
= {\la\pi^+K^-|(\bar{u}\lambda^a d)(\bar{s}
\lambda^a c)|D^0\ra \over \la\pi^+K^-|(\bar{u}d)(\bar{s}c)|D^0\ra_f}, \non \\
r_2(D\to\bar{K}\pi) &=& {\la\pi^+\bar{K}^0|(\bar{s}\lambda^a d)(\bar{u}
\lambda^a c)|D^+\ra \over \la\pi^+\bar{K}^0|(\bar{s}d)(\bar{u}c)|D^+\ra_f}
= {\la\pi^0\bar{K}^0|(\bar{s}\lambda^a d)(\bar{u}
\lambda^a c)|D^0\ra \over \la\pi^0\bar{K}^0|(\bar{s}d)(\bar{u}c)|D^0\ra_f},
\en
where we have applied Eq.(6) to derive (11) and isospin symmetry to relate
the $D^+\to\bar{K}^0\pi^+$ matrix elements of color octet currents to that
of $D^0\to K^-\pi^+,~\bar{K}^0\pi^0$. In the literature, factorization often
means that the parameters $a_1$ and $a_2$
(and hence $\xi_1$ and $\xi_2$) are universal,
\footnote{Note that $\xi_1=\xi_2=1/N_c$ corresponds to naive factorization,
while leading $1/N_c$ expansion leads to $\xi_1=\xi_2=0$.}
namely they are channel
independent in $D$ or $B$ decays. However, we see from Eq.(13) that {\it
a priori} there is no reason to expect that $r_1$ and $r_2$ are decay mode
independent.

    For the form factor $f_+^{DK}(q^2)$ in Eq.(11) we will use the average
value
\be
f_+^{DK}(0)=0.76\pm 0.02
\en
extracted from the recent measurements of
$D\to K\ell\bar{\nu}$ by CLEO II, E687 and E691 [14]. As for the form factor
$f_+^{D\pi}$, there are only two available experimental information.
An earlier measurement
of the Cabibbo-suppressed decay $D^0\to \pi^-\ell^+\nu$ by Mark III yields
$\left|{f_+^{D\pi}(0)/ f_+^{DK}(0)}\right|=\,1.0^{+0.6}_{-0.3}\pm 0.1$ [14,15],
while a very recent CLEO-II measurement of $D^+\to \pi^0\ell^+\nu$ gives
$\left|{f_+^{D\pi}(0)/ f_+^{DK}(0)}\right|=\,1.29\pm 0.21\pm 0.11$ [16].
Though the latter perfers a larger $f_+^{D\pi}(0)$ over $f_+^{DK}(0)$,
its error is still very large. Fortunately we can use the recent measurement
of $D^+\to\pi^+\pi^0$ by CLEO II [17] to fix $f_+^{D\pi}(0)$. The amplitude of
$D^+\to\pi^+\pi^0$ reads
\be
A(D^+\to\pi^+\pi^0)=\,{G_F\over 2\sqrt{2}}V_{ud}^*V_{cs}(a_1+a_2)(m_D^2-m_
\pi^2)f_\pi f_+^{D\pi}(m_\pi^2),
\en
where $a_1$ and $a_2$ are defined in the same manner as in (12) except that
in the present case $r_1(D\to\pi\pi)=r_2(D\to\pi\pi)$. Assuming a monopole
behavior for the form factor
\be
f_+(q^2)=\,{f_+(0)\over 1-(q^2/m_*^2)},
\en
where $m_*$ is the mass of the low-lying $1^-$ resonance that couples to
the weak current, and using the experimental value [17]
\be
{\cal B}(D^+\to\pi^+\pi^0)=\,(0.22\pm 0.05\pm 0.05)\%,
\en
as well as $\tau(D^+)=10.66\times 10^{-13}s$ [18], we find [19]
\be
[1+\xi(D\to\pi\pi)]f_+^{D\pi}(0)\approx 0.83\,.
\en
To proceed further we assume that $\xi_1(D\to\bar{K}\pi)=\xi_2(D\to\bar{K}
\pi)$ (which is at least valid in the SU(3) limit)$=\xi(D\to\pi\pi)$.
Substituting (14), (16), (18) into (15) we find
$\xi(D\to\bar{K}\pi)\approx 7\times 10^{-3}$, and hence
\be
r(D\to\bar{K}\pi)\approx -{2\over 3}\,.
\en
Evidently, the subleading $1/N_c$ factorizable contribution is almost
compensated by the nonfactorizable soft gluon effect, so that
$\xi\approx 0$. This explains why the leading $1/N_c$ expansion operates for
$D\to \bar{K}\pi$ decays.

We digress for a moment to make a remark on the $W$-exchange amplitude in the
decays $D^0\to K^-\pi^+,~\bar{K}^0\pi^0$, which is given by
\be
A(W-{\rm exchange}) &=& {G_F\over\sqrt{2}} V_{ud}^*V_{cs}\Big\{c_2\la K^-\pi
^+|(\bar{s}d)|0\ra\la 0|(\bar{u}c)|D^0\ra    \non \\
&+& c_1\la K^-\pi^+|(\bar{s}\lambda^a d)(\bar{u}\lambda^a c)|D^0\ra\Big\}.
\en
We see that although the first term vanishes in the SU(3) limit owing to the
conservation of vector current, the effect
of soft-gluon exchange could be important. Indeed, a general phenomenological
analysis of $D\to\bar{K}\pi$ data indicates that $W$-exchange is small
compared to external and internal $W$-emission amplitudes but it is not
negligible [20].

   We next determine the parameters $r_1$ and $r_2$ for $D\to VP$ decays,
whose amplitudes are of the form
\be
A(D\to VP)=\,M(\ep\cdot p_D),
\en
where $\ep_\mu$ is the polarization vector of the vector meson $V$, and
$p_D$ is the momentum of the charmed meson. We first consider the decay
$D^+\to\phi\pi^+$ which proceeds solely through internal $W$ emission:
\be
A(D^+\to\phi\pi^+) &=& {G_F\over\sqrt{2}}V_{us}^*V_{cs}a_2\la\phi|(\bar{s}s)
|0\ra\la\pi^+|(\bar{u}c)|D^+\ra   \non \\
&=& {G_F\over\sqrt{2}}V_{us}^*V_{cs}b'(\ep\cdot p_D),
\en
with
\be
b'=\,2a_2f_\phi m_\phi f_+^{D\pi}(m^2_\phi),
\en
where we have applied the relation $\la\phi|(\bar{s}s)|0\ra=f_\phi
m_\phi\ep_\mu$. From the measured branching ratio ${\cal B}(D^+\to\phi
\pi^+)=(6.0\pm 0.8)\times 10^{-3}$ [18], and the decay rate formula
\be
\Gamma(D^+\to\phi\pi^+)=\,\left|{G_F\over\sqrt{2}}V_{us}^*V_{cs}b'\right|^2
\left({(m_D^2-m_\phi^2-m_\pi^2)^2\over 4m^2_\phi}-m_\pi^2\right),
\en
we obtain
\be
|b'(D^+\to\phi\pi^+)|\approx 0.336\,{\rm GeV}^3.
\en
Comparing (25) with (23) leads to ($f_\phi=0.221$)
\be
\xi_2(D^+\to\phi\pi^+)\approx -0.12\,,~~~~~r_2(D^+\to\phi\pi^+)\approx -0.92\,.
\en
It is evident that the soft-gluon effect is larger than that in $D\to PP$
decay owing to the small relative momentum between the $VP$ final states.

   The decay $D^+\to\bar{K}^{*0}\pi^+$ receives both external and internal
$W$-emission contributions:
\be
A(D^+\to\bar{K}^{*0}\pi^+) &=&{G_F\over \sqrt{2}}V_{ud}^*V_{cs}\left(a_1\la\pi
^+|(\bar{u}d)|0\ra\la\bar{K}^{*0}|(\bar{s}c)|D^+\ra+a_2\la\bar{K}^{*0}|(
\bar{s}d)|0\ra\la\pi^+|(\bar{u}c)|D^+\ra\right)     \non \\
&=& {G_F\over\sqrt{2}}V_{ud}^*V_{cs}(a'+b')(\ep\cdot p_D),
\en
with
\be
a' &=& a_1f_\pi[(m_D+m_{K^*})A_1^{DK^*}(m_\pi^2)-(m_D-m_{K^*})A_2^{DK^*}
(m_\pi^2)],   \non \\     b' &=& 2a_2f_{K^*}m_{K^*}f_+^{D\pi}(m^2_{K^*}),
\en
where use has been made of
\be
\la V(p_V)|(\bar{s}c)|D(p_D)\ra &=&\,i\Big\{\ep_\mu(m_D+m_{V})A_1(q^2)-
\ep\cdot q{(p_D+p_{V})_\mu\over m_D+m_{V}}A_2(q^2)  \\
&-& 2{\ep\cdot q\over q^2}q_\mu m_{V}[A_3(q^2)-A_0(q^2)]\Big\}+{2\over
m_D+m_{V}}\epsilon_{\mu\nu\alpha\beta}\ep^\nu p^\alpha_D p^\beta_{V}
V(q^2),  \non
\en
with $q_\mu=(p_D-p_{V})_\mu$ and $A_3(0)=A_0(0)$. Assuming $\xi_1\approx
\xi_2$ and using the experimental branching
ratio ${\cal B}(D^+\to\bar{K}^{*0}\pi^+)=(1.9\pm 0.7)\%$ [18], and the
measured form factors [14]
\be
A_1^{DK^*}(0)=\,0.59\pm 0.04\,,~~~~A_2^{DK^*}(0)=\,0.44\pm 0.09\,,
\en
and $f_{K^*}=0.220$ determined from the decay $\tau\to K^{*-}\nu_\tau$, we get
\footnote{For some unknown reasons, the result for $\xi$ determined from the
decay $D^+\to\bar{K}^0\rho^+$ is very unsatisfactory: $\xi(D\to \bar{K}\rho)
\sim -1.4$.}
\be
\xi(D\to\bar{K}^*\pi)\approx -0.22\,,~~~~r(D\to\bar{K}^*\pi)\approx -1.10\,.
\en

    Thus far we have assumed $\xi_1=\xi_2$ or $r_1=r_2$. Whether this is
true or not can be tested from the decay $D^+_s\to\phi\pi^+$ which occurs
solely through the external $W$-emission diagram
\be
A(D^+_s\to\phi\pi^+)=\,{G_F\over\sqrt{2}}V_{ud}^*V_{cs}a'(\ep\cdot p_{_{
D_s}}),
\en
with
\be
a'=\,a_1f_\pi[(m_{D_s}+m_\phi)A_1^{D_s\phi}(m_\pi^2)-(m_{D_s}-m_\phi)A_2^{
D_s\phi}(m_\pi^2)].
\en
{}From ${\cal B}(D_s^+\to\phi\pi^+)=3.6\%$ [21] and $\tau(D_s^+)=4.50\times
10^{-13}s$ [18], we find
\be
|a'|=\, 0.25\,{\rm GeV}^3.
\en
If the form factors
$A_{1,2}^{D_s\phi}$ are taken to be the same as $A_{1,2}^{DK^*}$, we then
obtain
\be
\xi_1(D_s\to\phi\pi)\approx -0.27\,,~~~~~r_1(D_s\to\phi\pi)\approx -1.22\,.
\en
Since $\xi_{1,2}=(a_{1,2}-c_{1,2})/c_{2,1}$ and $|c_2|<<|c_1|$, it is clear
that the determination of $\xi_2$ is far more less uncertain than $\xi_1$.
Within uncertainties we see that $r_1$ and $r_2$ are very close. Note that
internal $W$ emission is usually suppressed relative to external $W$ emission
owing to color mismatch. However, we see from (25) and (34) that internal $W$
emission in $D\to\phi\pi,~D\to\bar{K}\pi$ dominates over external $W$
emission. Of course, nonfactorizable soft-gluon contribution plays an
essential role here.

   For the decay $D\to VV$, its general amplitude reads
\be
A(D\to V_1V_2)=i{G_F\over\sqrt{2}}V_{cq_{_1}}^*V_{q_{_2}q_{_3}}\ep^\mu(V_1)
\ep^\nu(V_2)(\hat{A}_1g_{\mu\nu}+\hat{A}_2p_\mu^D p_\nu^D+i\hat{V}\epsilon_{
\mu\nu\alpha\beta}p^\alpha_Dp_1^\beta).
\en
The first term is an $S$-wave amplitude, the second is a longitudinal
$D$-wave term, and the third is a $P$-wave term. We find for $D^+\to\bar{K}^
{*0}\rho^+$ decay
\be
\hat{A}_1 &=& a_1f_\rho m_\rho(m_D+m_{K^*})A_1^{DK^*}(m_\rho^2)+a_2f_{K^*}
m_{K^*}(m_D+m_\rho)A_1^{D\rho}(m^2_{K^*}),   \non \\
\hat{A}_2 &=& -{2\over m_D+m_{K^*}}a_1f_\rho m_\rho A_2^{DK^*}(m^2_\rho)-{
2\over m_D+m_\rho}a_2f_{K^*}m_{K^*}A_2^{D\rho}(m^2_{K^*}), \\
\hat{V} &=& -{2\over m_D+m_{K^*}}a_1f_\rho m_\rho V^{DK^*}(m^2_\rho)-{
2\over m_D+m_\rho}a_2f_{K^*}m_{K^*}V^{D\rho}(m^2_{K^*}), \non
\en
where we have applied Eq.(29).
Experimentally, a mixture of transverse and longitudinal polarization is
found to be consistent with a pure $S$-wave term [22,18]:
\be
&& {\cal B}(D^+\to\bar{K}^{*0}\pi^+;~S-{\rm wave})=\,\left(4.1^{+1.5}_{-1.2}
\right)\%,   \non \\
&& {\cal B}(D^+\to\bar{K}^{*0}\pi^+;~P-{\rm wave})<5\times 10^{-3},  \\
&&{\cal B}(D^+\to\bar{K}^{*0}\pi^+;~D-{\rm wave~longitudinal})<7\times
10^{-3}. \non
\en
It is convenient to write the Lorentz invariant amplitude in terms of three
helicity amplitudes:
\be
H_{00} &=& {1\over m_{K^*}m_\rho}\left[{1\over 2}(m_D^2-m^2_{K^*}-m^2_\rho)
\hat{A}_1+m^2_Dp_c^2\hat{A}_2\right], \non \\
H_{++} &=& \hat{A}_1+m_Dp_c\hat{V},   \\
H_{--} &=& \hat{A}_1-m_Dp_c\hat{V},  \non
\en
where $p_c$ is the c.m. momentum. The decay rate is then given by
\be
\Gamma(D^+\to\bar{K}^{*0}\rho^+)=\,{p_c\over 8\pi m_D^2}(|H_{00}|^2+|H_{++}|^2
+|H_{--}|^2).
\en
Assuming $r_1=r_2$ as before, $A_1^{D\rho}(0)/A_1^{DK^*}(0)=0.89$ [23], and a
monopole behavior for form factors [$m_*=2.53$ GeV for $(\bar{s}c)$ current
and 2.42 GeV for $(\bar{u}c)$ current], we obtain
\be
\xi(D\to \bar{K}^*\rho)\approx -0.34\,,~~~~r(D\to\bar{K}^*\rho)\approx -1.35\,.
\en
The value of $\xi$ is sensitive to the ratio $A_1^{D\rho}(0)/
A_1^{DK^*}(0)$ which has not been measured. Nevertheless, it does not affect
the general feature that $|r(D\to VV)|~{\large ^>_{\sim}}~|r(D\to VP)|$.

    To conclude this section, the parameter $r_2$, which measures the
nonfactorizable contribution from color octet currents, is found to be of
order $\sim -2/3,~-(0.9\sim 1.1),~-(1.2\sim 1.3)$ respectively for
$D\to PP,~VP,
{}~VV$ decays, in accordance with the expectation that soft-gluon effects
become
stronger when the final-state particles become less energetic. The parameters
$a_1$ and especially $a_2$ are thus not universal; they are channel
or class dependent.
\vskip 0.7 cm
3.~~{\it Bottom Meson Decays}~~~In the exclusive two-body nonleptonic weak
decays of bottom mesons, the parameter $a_1$ can be estimated directly
from neutral $B$ decays, e.g. $\bar{B}^0\to D^+\pi^-,~D^+\rho^-$. This
comes from the fact that, contrary to the charmed meson case, both $W$
exchange and final-state interactions are presumably negligible as the
decay particles are very energetic (for a recent study of final-state
interactions in $B$ decays, see Ref.[24]).
As for the parameter $a_2$, $\bar{B}\to
\psi\bar{K}^{(*)}$ are thus far the only color-suppressed modes which
have been measured experimentally. Hence, a direct measurement of $a_2$
is available only in these decay modes. A fit to the CLEO II data of
${B}^-\to\psi K^{-(*)},~\bar{B}^0\to\psi\bar{K}^{0(*)}$ yields [8] (see also
Ref.[25])
\be
|a_2(\bar{B}\to\psi\bar{K})|=\,0.26\pm 0.02\,.
\en
Since the Wilson coefficients at the renormalization scale $\mu\sim m_b$
are given by
\be
c_1(m_b)\sim 1.11\,,~~~~c_2(m_b)\sim -0.26\,,
\en
it follows two possibilities: either $a_2<0$ and
\be
\xi_2(\bar{B}\to\psi\bar{K})\approx 0\,,~~~~r_2(\bar{B}\to\psi\bar{K})\approx
-{2\over 3},
\en
which is precisely what expected from the leading $1/N_c$ expansion, or
$a_2>0$ and
\be
\xi_2(\bar{B}\to\psi\bar{K})\approx 0.47\,,~~~~r_2(\bar{B}\to\psi\bar{K})
\approx 0.27\,.
\en
Unfortunately, what is the sign of $a_2$ for $\bar{B}\to\psi\bar{K}^{(*)}$
decays is still very confusing. On the one hand, a recent calculation based on
QCD sum rule indicates a destructive interference between the nonfactorizable
soft-gluon term and the subleading $1/N_c$ factorizable term [13]. This in
turn implies a negative $a_2$. On the other hand, the ratio $a_2/a_1$ is
found to be positive in ${B}^-\to D\pi,~D\rho,~D^*\pi,~D^*\rho$ decays
[8].
\footnote{The ratio $a_2/a_1=0.23\pm 0.11$ obtained in Ref.[8] actually
comes from (i) the determination of $|a_1|=1.15\pm 0.11$ from $\bar{B}^0\to
D^{(*)}\pi (\rho)$ decays, (ii) $|a_2|=0.26\pm 0.02$ from
$\bar{B}\to\psi\bar{K}
^{(*)}$ decays, and (iii) the constructive interference in $B^-\to D^{(*)}\pi
(\rho)$ decays. {\it A priori} there is no reason to expect that $|a_2(\bar{B}
\to D^{(*)}\pi(\rho))|=|a_2(\bar{B}\to\psi\bar{K}^{(*)})|$ as the c.m.
momentum in $\bar{B}\to\psi\bar{K}$ is 1683 MeV, while it is 2307 MeV in
$\bar{B}\to D\pi$ decay.}

In the following, we will argue that $a_2$ appears to be channel dependent:
Its sign is opposite in $\bar{B}\to\psi\bar{K}^{(*)}$ and
$\bar{B}\to D^{(*)}\pi(\rho)$ decays. Following the argument presented in
the last section, it is expected that
\be
|r_2(\bar{B}\to\psi\bar{K})|>|r_2(\bar{B}\to D\pi)|.
\en
A fit of the theoretical calculation [26]
\be
R_1=\,{{\cal B}(B^-\to D^0\pi^-)\over {\cal B}(\bar{B}^0\to D^+\pi^-)}=\left(
1+1.23{a_2\over a_1}\right)^2
\en
to the measured value $R_1=1.89\pm 0.26\pm 0.32$ [8] gives
\footnote{A direct fit of $a_2/a_1$ to the other ratios $R_2,~R_3,~R_4$ (for a
definition, see Ref.[26]) also gives a value substantially larger than 0.23
(see Eqs.(27-30) of Ref.[8]).}
\be
{a_2\over a_1}(\bar{B}\to D\pi)\approx 0.30\,.
\en
When combining with $a_1\approx 1.05$ determined from $\bar{B}^0\to D^+\pi^-$,
this leads to
\footnote{Unlike the charmed meson decay, $r_2$ here is supposed not to vary
significantly from $\bar{B}\to D\pi$ to $\bar{B}\to D^*\rho$ decays.}
\be
a_2(\bar{B}\to D\pi)\approx 0.32\,,~~~\xi_2(\bar{B}\to D\pi)\approx 0.52\,,~~~
r_2(\bar{B}\to D\pi)\approx 0.36\,.
\en
Evidently, the relation (46) can be satisfied only if $a_2(\bar{B}\to\psi
\bar{K})$ is negative:
\be
a_2(\bar{B}\to\psi\bar{K})=\,-(0.26\pm 0.02).
\en
This argument that the nonfactorizable term and the subleading $1/N_c$
factorizable term in $\bar{B}\to\psi\bar{K}$ decay are in opposite signs is
also in accordance with a recent QCD sum rule calculation [13].

    The most striking feature observed by CLEO [8] is that the parameter
$r_2(\bar{B}\to D^{(*)}\pi(\rho))$ is positive, recalling that the analogous
quantity is always negative in charm decay. If our above argument is correct,
we have to understand why the sign of $r_2$ changes dramatically from
$\bar{B}\to\psi
\bar{K}$ to $\bar{B}\to D^{(*)}\pi(\rho)$ decays. This should be checked by
lattice and QCD sum rule calculations. We see in charmed meson decay that the
parameters $r_1$ and $r_2$ are empirically close. Is this still true in
bottom decay? The QCD sum rule calculation in Ref.[12] indicates
$r_1(\bar{B}^0\to D^+\pi^-)\approx -1$
\footnote{This number is taken from Ref.[27], which is smaller
than the original estimate $r_1\sim -1.5$ given in Ref.[12].}
with a sign opposite to $r_2(B^-\to D\pi)$. However, a direct fit to the
data of $\bar{B}^0\to D^{(*)}\pi(\rho)$ (except for
$\bar{B}^0\to D^+\rho^-$) gives a value of $a_1$ of order $1.01\sim 1.05$ (see
Ref.[24] and Table XX of Ref.[8]) with the neglect of $W$ exchange and
final-state interactions. It thus appears that the sign of
$r_1(\bar{B}^0\to D^+\pi^-)$ could be the same as that of $r_2(B^-\to D\pi)$.
This issue should be clarified in the future.

\vskip 0.8 cm

    4.~~{\it Conclusions}~~~We have considered the nonfactorizable
contributions to exclusive two-body nonleptonic weak decays of charmed and
bottom mesons arising from color octet currents. From the data we have
determined the parameter $r_2$, which measures the soft gluon effects, to be
$\sim -0.67\,$, $-(0.9\sim 0.11)$, $-(1.2\sim 1.3)$, $\sim 0.36$
respectively for $D\to \bar{K}\pi,~\bar{K}^*\pi,~\bar{K}^*\rho$, and
$\bar{B}\to D\pi$ decays. Therefore,
the soft-gluon effects become more important when final-state particles become
less energetic, as expected. In the $1/N_c$ approach, the subleading
contribution is characterized by the parameters $\xi_i=(1/N_c+r_i/2)\,
(i=1,2)$.
It is evident that the coefficients $a_1\,(=c_1+\xi_1c_2)$ and especially
$a_2\,(=c_2+\xi_2c_1)$ are not universal;
they are decay mode or class dependent.

   The very striking feature observed by CLEO that $r_2$ is positive for
$\bar{B}\to D\pi$ decays, whereas it is always negative in charmed meson
decays, needs to be checked
by lattice and QCD sum rule calculations. We have argued that for
$\bar{B}\to\psi\bar{K}^{(*)}$ decays, the nonfactorizable term contributes
destructively with the $1/N_c$ factorizable term so that
the sign of $a_2$ is negative. As a consequence, the
leading $1/N_c$ expansion turns out working well for
$\bar{B}\to\psi\bar{K}^{(*)}$, though it fails for $\bar{B}\to D^{(*)}\pi
(\rho)$ decays.

\vskip 2.0 cm
\centerline{\bf ACKNOWLEDGMENT}

    This work was supported in part by the National Science Council of ROC
under Contract No. NSC83-0208-M-001-014.

\vskip 1.7 cm
\centerline{\bf References}
\medskip
\begin{enumerate}

\item D. Gross, talk presented at the XVI International Symposium
on Lepton-Photon Interactions, Ithaca, 10-15 August 1993.

\item H. Georgi, {\sl Annu. Rev. Nucl. Part. Sci.} {\bf 43}, 209 (1993).

\item R. Dashen and A.V. Manohar, \pl {\bf B315}, 425, 438 (1993); R. Dashen,
E. Jenkins, and A.V. Manohar, \pr {\bf D49}, 4713 (1994).

\item A.J. Buras, J.-M. G\'erard, and R. R\"uckl, \np {\bf B268}, 16 (1986).

\item H.Y. Cheng, {\sl Int. J. Mod. Phys.} {\bf A4}, 495
(1989).

\item M. Fukugita,
T. Inami, N. Sakai, and S. Yazaki, \pl {\bf 72B}, 237 (1977); D. Tadi\'c and
J. Trampeti\'c, \pl {\bf 114B}, 179 (1982).

\item M. Bauer, B. Stech, and M. Wirbel, \zp {\bf C34}, 103 (1987).

\item CLEO Collaboration, M.S. Alam {\it et al.,} CLNS 84-1270 (1994).

\item H.Y. Cheng, \zp {\bf C32}, 237 (1986).

\item N. Deshpande, M. Gronau, and D. Sutherland, \pl {\bf 90B}, 431 (1980);
M. Gronau and D. Sutherland, \np {\bf B183}, 367 (1981).

\item B. Blok and M. Shifman, {\sl Sov. J. Nucl. Phys.} {\bf 45}, 35, 301,
522 (1987).

\item B. Blok and M. Shifman, \np {\bf B389}, 534 (1993).

\item A. Khodjamirian and R. R\"uckl, MPI-PhT/94-26 (1994).

\item M. Witherell, talk presented at the XVI International Symposium
on Lepton-Photon Interactions, Ithaca, 10-15 August 1993.

\item Mark III Collaboration, J. Adler {\it et al.,} \prl {\bf 62}, 1821
(1989).

\item CLEO Collaboration, M.S. Alam {\it et al.,} \prl {\bf 71}, 1311 (1993).

\item CLEO Collaboration, M. Selen {\it et al.,} \prl {\bf 71}, 1973 (1993).

\item Particle Data Group, \pr {\bf D45}, S1 (1992).

\item L.L. Chau and H.Y. Cheng, ITP-SB-93-49, to appear in {\sl Phys. Lett.}
{\bf B}.

\item L.L. Chau and H.Y. Cheng, \prl {\bf 56}, 1655 (1986); \pr {\bf D36},
137 (1987); \pl {\bf B222}, 285 (1989).

\item F. Muheim and S. Stone, \pr {\bf D49}, 3767 (1994).

\item Mark III Collaboration, D. Coffman {\it et al.,} \pr {\bf D45}, 2196
(1992).

\item M. Wirbel, B. Stech, and M. Bauer, \zp {\bf C29}, 637 (1985).

\item H. Yamamoto, HUTP-94/A006 (1994).

\item A. Deandrea, N. Di Bartolomeo, R. Gatto, and G. Nardulli, \pl {\bf
B318}, 549 (1993); K. Honscheid, K.R. Schubert, and R. Waldi,
OHSTPY-HEP-E-93-017 (1993); S. Resag and M. Beyer, BONN-TK-93-18 (1993);
M. Gourdin, A.N. Kamal, Y.Y. Keun, and X.Y. Pham,
PAR/LPTHE/94-22 (1994).

\item M. Neubert, V. Riekert, Q.P. Xu, and B. Stech, in {\it Heavy Flavors},
edited by A.J. Buras and H. Lindner (World Scientific, Singapore, 1992).

\item I. Bigi, B. Blok, M. Shifman, N. Uraltsev, and A. Vainshtein,
CERN-TH.7132/94 (1994).

\end{enumerate}

\end{document}